# Raman spectroscopy study of $Na_xCoO_2$ and superconducting $Na_xCoO_2.yH_2O$

Y.G. Shi, [1] Y.L. Liu, [1] H.X. Yang, [1]* C. J. Nie, [1] R. Jin [2] and J.Q. Li[1]

1. Institute of Physics, Chinese Academy of Sciences, Beijing, People's Republic of China
2. Condensed Matter Sciences Division, Oak Ridge National Laboratory, Oak Ridge, Tennessee 37831, USA

*author for correspondence: hxyang@blem.ac.cn

The Raman spectra of the parent compound $Na_xCoO_2$ (x=0.75) and the superconducting oxyhydrates $Na_xCoO_2 \cdot yH_2O$ with different superconducting temperatures ($T_c$) have been measured. Five Raman active phonons around 195 cm$^{-1}$ ($E_{1g}$), 482 cm$^{-1}$, 522 cm$^{-1}$, 616 cm$^{-1}$ ($3E_{2g}$), 663 cm$^{-1}$ ($A_{1g}$) appear in all spectra. These spectra change systematically along with the intercalation of $H_2O$ and superconducting properties. In particular, the Raman active phonons ($A_{1g}$ and $E_{1g}$) involving the oxygen motions within the Co-O layers show up monotonous decrease in frequency along with superconducting temperature $T_c$. The fundamental properties and alternations of other active Raman phonons in the superconducting materials have also been discussed.

Understanding the physical mechanism of the well-known high-transition-temperature superconducting cuprates has been a longstanding challenge since its discovery in 1986[1]. The recent discovery of new superconducting system, layered sodium cobalt oxyhydrate $Na_{0.3}CoO_2$•$1.3H_2O$ with $T_c$≈5K [2], might help shed light on this issue due to its remarkable resemblance in both structure and superconducting properties to the cuprates. In order to obtain deeper insight into physical mechanism, systematically theoretical and experimental investigations on this new superconducting system are highly desirable to gain detailed knowledge for facilitating the comparison. In our previous paper, fundamental structural features and physical properties of $Na_xCoO_2$ (0.5<x<1.0) and the hydrated $Na_xCoO_2 \cdot yH_2O$ superconducting materials have been investigated [3, 4]. In the present work, we report a comprehensive study of Raman scattering on $Na_xCoO_2 \cdot 1.3H_2O$ oxyhydrates and its unhydrated analogs. Actually, Raman scattering has been proved to be a valuable technique for probing the structural transformation between normal and the superconducting states as used in the studies of unconventional superconductors, such as $YBa_2Cu_3O_y$[5, 6, 7, 8], bismuthates [9], the alkalidoped fullerides [10] and the organic superconductors [11, 12].

Ceramic pellets, single crystal $Na_xCoO_2$(x=0.75) and a series of $Na_xCoO_2 \cdot 1.3H_2O$ oxyhydrates with different superconducting critical temperatures were used in the present Raman scattering experiments. $Na_xCoO_2$ ceramic pellets were synthesized following a procedure as described Ref.13. High-quality $Na_xCoO_2$ single crystals with size of around 3×3×0.2mm were grown using the flux method [14]. Superconducting $Na_xCoO_2 \cdot 1.3H_2O$ materials were synthesized from the ceramic pellets $Na_xCoO_2$ through a chemical oxidation process as reported previously [3, 15]. The magnetic susceptibility of $Na_xCoO_2 \cdot 1.3H_2O$ samples was measured by zero-field cooling (ZFC) DC in a field of 20Oe. Raman spectra were collected in back-scattering geometry at room temperature using a Jobin-Yvon T64000 triple spectrometer equipped with a cooled change-couple device. In the spectrometer an objective of 100X-magnification was used to focus the laser beam on the sample surface and to collect the scattered light. Two excitation wavelengths at 488.0nm and 514.5 nm of an $Ar^+$ ion laser were used in our experiments. The laser power at the focus spot of 2-3 μm in diameter was kept below 1 mW to prevent laser-induce damage to the samples.

Fig. 1(a) shows the results of magnetic susceptibility as a function of temperature for a series of $Na_xCoO_2 \cdot 1.3H_2O$ oxyhydrates. The diamagnetic signals appear in all samples below *Tc* provide direct evidence for bulk superconductivity. The critical superconducting temperature of the samples was further analyzed in correlation with the c-axis lattice parameters of the samples as calculated from XRD data. Which reveals an evident inverse correlation between the $CoO_2$-layer distance and the superconducting transition temperature in $Na_xCoO_2 \cdot 1.3H_2O$ oxyhydrates (Fig. 1(b)).

In order to identify the phonon modes of $Na_xCoO_2$, a single crystalline sample of typical size 3×3×0.2mm was used to measure Raman polarized spectra. The single crystal consists of hexagonal thin platelets as shown in the image of Fig. 2(a). A visual observation under microscope reveals that each crystal platelet is in fact a pack of thinner layers (d < 5 μm) parallel to the hexagon surfaces. All diffraction peaks in the corresponding XRD pattern can be indexed by a hexagonal cell with lattice parameters a =2.834Å, and c=10.94Å. The a-axis parameter agrees with the data for the ceramic samples as reported in our previous paper [3], while the c-axis parameter is slightly longer.

The Raman spectra are very sensitive to properties of the sample surface. Fig. 2(b) shows a typical result obtained from the as-grown sample without pre-cleaning. It is noted that this spectrum shows up certain similarities with those reported in Ref. 16. The EDAX analysis (see the inset of Fig. 2(b)) suggests that the surface of the as-made sample is covered by impurity phases mainly identified as $Na_2CO_3$. The formation of $Na_2CO_3$ on the sample surface can be explained as following; when the $Na_xCoO_2$ compound is stored under ambient condition, it reacts with water in air, decomposes into $Na_2O$, and then further react with the $CO_2$ to form a more stable phase $Na_2CO_3$ . On the other hand, the migration of $Na^+$ ions in this kind of materials also contributes to this phenomenon. As Ronald et al discussed in Ref. 17, the structure of $Na_xCoO_2$ consists of $CoO_6$ sheets and $Na^+$ ions intercalated within a trigonal prismatic site between the $CoO_6$ sheets. The distance between the faces of oxygen ions above and below the sodium is 0.81Å, which is wide enough to allow the sodium ion moving freely through this material, as there is a tunnel available for motion between oxygen sheets. Hence, it is indeed necessary to clean the surface of $Na_xCoO_2$ samples thoroughly before Raman scattering measurements to ensure precise data to be obtained.

The hexagonal crystal structure of $Na_xCoO_2$ consisting of $CoO_2$ and Na layers parallel to the ab planes belongs to space group ***$D_{6h}$ ( $P6_3$/mmc, Z=2 )***. Its vibrations at a wave vector of $q \cong 0$ are classified following the irreducible representation of the factor group. There are five Raman active phonon modes: $A_{1g} + E_{1g} + 3E_{2g}$, which can be identified unambiguously by specific polarization configurations. Their second-order susceptibilities are restricted by the symmetric properties:

$$\chi(A_{1g}) = \begin{bmatrix} a & & \\ & a & \\ & & b \end{bmatrix} \quad \chi(E_{1g}) = \begin{bmatrix} & & \\ & & d \\ & e & \end{bmatrix} \begin{bmatrix} & & -d \\ & & \\ -c & & \end{bmatrix} \quad \chi(E_{2g}) = \begin{bmatrix} & f & \\ f & & \\ & & \end{bmatrix} \begin{bmatrix} f & & \\ & -f & \\ & & \end{bmatrix}$$

where the absence of an entry in the matrix position ***ij*** implies a zero component, the letters in the matrices indicate the nonzero components, and the occurrence of the same letter in different positions indicates equal components. The cross section of Raman scattering is proportional to

$$\left| e_S^i \chi^{ij} e_S^j \right|^2$$

where the superscripts refer to Cartesian components *x*, *y* and *z* or the position of matrix components, subscripts ***S*** is for scattering and ***I*** for incident, and ***e*** is the unit parallel to the electric field.

Polarized Raman spectra were measured from ***ab*** and ***ac*** single crystal surfaces of $Na_xCoO_2$. In order to insure that the incident and scattered polarizations are parallel to the crystalline, the following precautions were taken: (i) Because the light is incident normal to the crystal surface, its polarization was made to be vertical so that it can be parallel to the [100]or [010] axis; and (ii) the direction of the scattered light was chosen to be normal to the crystal surface so that the scattered polarization V or H can be parallel to the [100] or [010]zone-axis direction. We have calculated the relative scattering intensities associated with $A_{1g}$, $E_{1g}$, $3E_{2g}$ phonons, and listed in Table I (A) and I (B) for the different crystal orientations and polarizations. Five modes ($A_{1g}$+$E_{1g}$+3$E_{2g}$) are Raman active. The $A_{1g}$ and $E_{1g}$ modes involve motions of the only oxygen atoms. Both Na and O motions may participate in the $E_{2g}$ modes. Co motions are not Raman active.

Fig. 3(a) and (b) display the Raman spectra from a single crystal sample from the ab plane and a surface parallel to the c axis direction. From these measurements in addition with above analyses for the $Na_xCoO_2$ materials with space group of ***P6₃/mmc*** and point group of ***6/mmm***(O6h), the all Raman active modes can be identified respectively as $A_{1g}$ at 663.81cm$^{-1}$, $E_{1g}$ at 195.8cm$^{-1}$, $E_{2g}$ at 482.13cm$^{-1}$, 522.35cm$^{-1}$, and 616.93 cm$^{-1}$.

Fig. 4(a) shows the Raman spectra for a parent and three superconducting samples.

Before offset for clarify, all the measured spectra have a constant intensity background; similar behavior has also been observed in other families of high $T_c$ conductors and this phenomenon has been attributed to the scattering by charge carries [12]. In each spectrum five clear modes can be evidently recognized, the data from the $Na_xCoO_2$ (x=0.7) material shows up noticeable similarities in comparison with results shown in Fig. 3 obtained from the single crystal sample. Because the crystal structure of ceramic $Na_xCoO_2$ and $Na_xCoO_2 \cdot 1.3H_2O$ materials have the crystal lattices with the same symmetric property, the Raman active modes for the superconducting samples have the fundamental similar features as clearly illustrated in Fig. 4(a), no additional Raman modes caused by $H_2O$ molecules intercalated between $CoO_2$ layers were observed. It is noted that the Raman active modes appearing in the spectra of Fig. 4(a) show systematic alternations along the changes of superconductivity. In Fig. 4(b), we have used dished lines to clearly indicate the shifting of the $A_{1g}$ and $E_{1g}$ modes. It is remarkable that $A_{1g}$ and $E_{1g}$ peaks of the superconducting samples show significant shifts towards lower frequencies in comparison with the parent sample, for example, the $A_{1g}$ peak appearing at 673cm$^{-1}$ in the parent material shifts to 663cm$^{-1}$ in the superconducting sample with $T_c$ =4.2K. Measurements on the superconducting samples also indicated that all five peaks shift monotonously towards the lower frequencies with the increase of $T_c$, and the $A_{1g}$ and $E_{1g}$ peaks in general have much larger shifts than the three $E_{2g}$ modes. Moreover, detailed examinations of the spectra also reveals that the full width at half maximum (FWHM) of the $A_{1g}$ peak becomes progressively shorter with the increase of $T_c$ in superconducting samples, and the FWHM of the parent material is much smaller than that of the superconducting samples. As pointed out in above context, $A_{1g}$ and $E_{1g}$ modes are essentially in connection with the movement of oxygen atoms within the $CoO_2$ layers. Hence, our measurements directly demonstrate the modification of oxygen motion in the crystal lattice arising from oxyhydrating plays a key role for understanding the superconductivity occurring in present system.

Analyses in combination with the data showed in Fig. 2(b) reveal a strong frequency dependence of the observed lattice modes on the c-axis parameter for the superconducting samples; the Raman peaks shift towards lower frequencies with the decrease of c-axis parameter, while $T_c$ is increasing . On the other hand, it is also known that the oxyhydrated superconducting phases have a much longer c-axis than the parent material, the Raman peaks are found to shift evidently towards lower frequency with the intercalation of $H_2O$. However, it is believed that the modifications of the phonons in these two cases might have evidently different underlying structural and physical origins.

It is worthy to point out that our results can be used to complete certain conclusions drawn by M. Dressel on the correlation of phonon frequencies with the superconducting transition temperatures of some (BEDT-TTF) salts, which suggests that, in high $T_c$ materials, the higher phonon frequencies possibly contribute to the increase of $T_c$ in certain systems, however, the issues on the connection of phonon and superconductivity in general is much more complex than that described by a BCS expression [18]. Therefore, in present case, it is also possible that a higher $T_c$ value doesn't have to in connection with soften lattice, but it is still possible the 'volume effect' [18], showing up as changes in the c-axis parameter, maybe operate in the $Na_xCoO_2 \cdot 1.3H_2O$ system.

In summery, we have investigated the Raman spectra of the single crystal, ceramic pellets of $Na_xCoO_2$(x=0.7) and superconducting $Na_xCoO_2 \cdot 1.3H_2O$ materials. We have identified that five active phonons generally appear in the Raman spectra of this kind of materials at the position of 663 $cm^{-1}$ ($A_{1g}$), 195 $cm^{-1}$ ($E_{1g}$) and 482 $cm^{-1}$, 522 $cm^{-1}$, 616 $cm^{-1}$($3E_{2g}$). The Raman active phonons, in particular the $A_{1g}$ and $E_{1g}$ modes and the FWHM change systematically along with the intercalation of $H_2O$ and the modification of superconducting properties.

---

## Acknowledgments

We would like to thank Professor N.L. Wang for providing single crystal of $Na_xCoO_2$ and Miss G. Zhu for the assistance in preparing samples and measuring Raman spectra. The work reported here is supported by the 'Outstanding Youth Fund' organized by National Natural Foundation of China.

**Table 1: Relative phonon Raman intensities for crystal orientations shown in Fig. 3 (a) and (b) for $Na_xCoO_2$, VV and VH refer to the different polarization combinations for the incident and scattered light.**

| ***Phonon Symmetry*** | ***ab-plane*** | | ***ac-plane*** | |
|---|---|---|---|---|
| | VV | VH | VV | VH |
| $A_{1g}$ | $a^2$ | $0$ | $b^2$ | $0$ |
| $E_{1g}$ | $0$ | $0$ | $0$ | $e^2$ |
| $E_{2g}$ | $f^2$ | $f^2$ | $0$ | $0$ |

**Table 2: Full width at half maximum (FWHM) vs. the superconducting critical temperature $T_c$**

| ***Tc*** | Parent material | 3.5K | 3.9K | 4.2K |
|---|---|---|---|---|
| **FWHM ($cm^{-1}$)** | 18 | 25 | 23 | 22 |

# Figure Captions

**Fig. 1**: (a) magnetic susceptibilities for $Na_xCoO_2 \cdot yH_2O$ samples showing the superconducting transitions at different temperatures.

(b) Transition temperatures ($T_c$) vs. c-axis parameter in $Na_xCoO_2 \cdot yH_2O$ superconductors.

**Fig. 2**: (a) SEM images showing the microstructure features of a $Na_xCoO_2$ single crystal. (b) Raman spectra from a $Na_xCoO_2$ single crystal sample with $Na_2CO_3$ on the surface, inset shows the data of EDAX analysis.

**Fig. 3**: (a) Raman spectra taken from the ab-plane of a $Na_xCoO_2$ single crystal. (b) Raman spectra taken from the surface parallel to the c axis direction.

**Fig. 4**: (a) Raman spectra of the parent and three superconducting samples. (b) $A_{1g}$ and $E_{1g}$ modes of the parent and the superconducting samples, the shifts of these modes are clearly indicated.

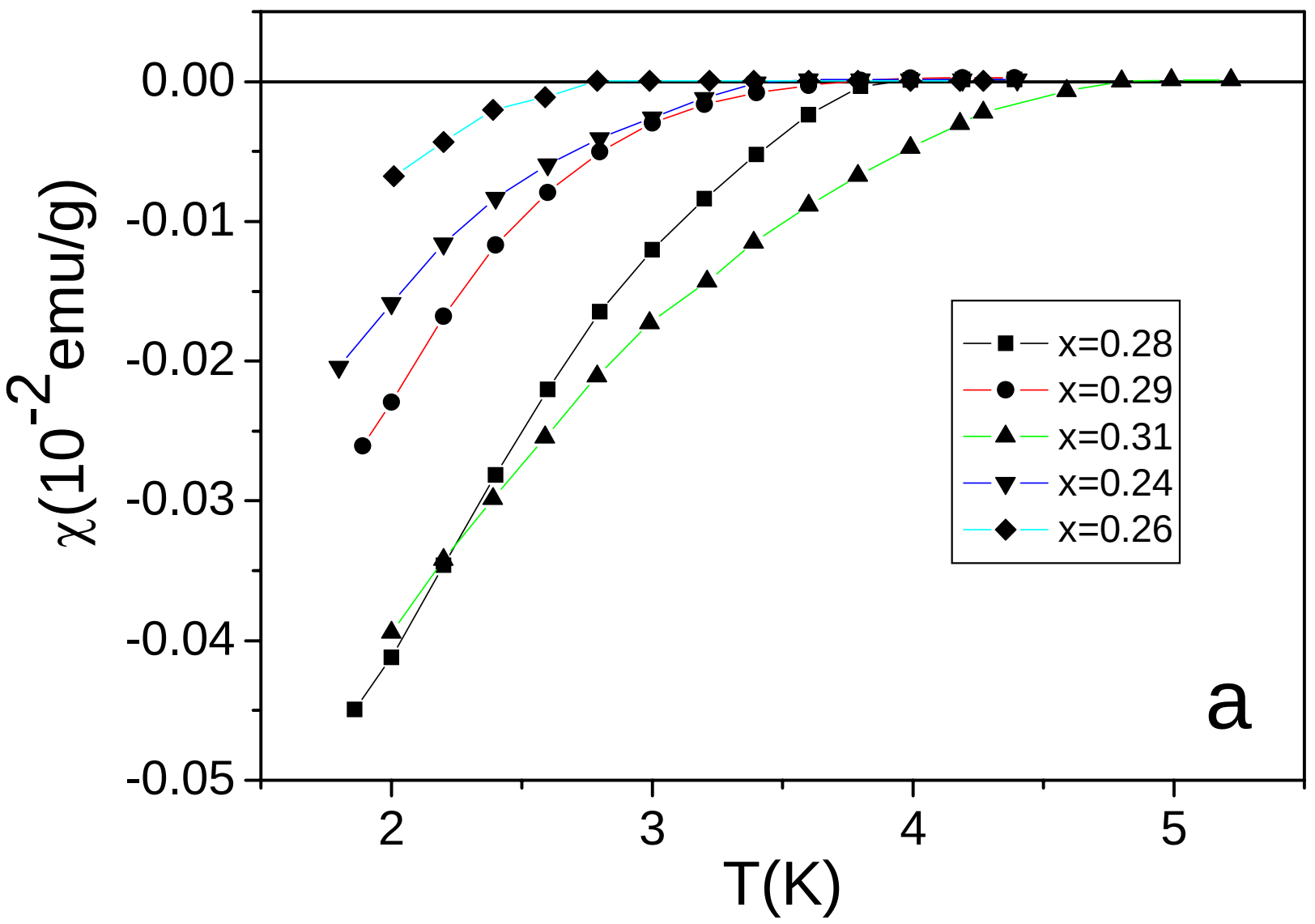

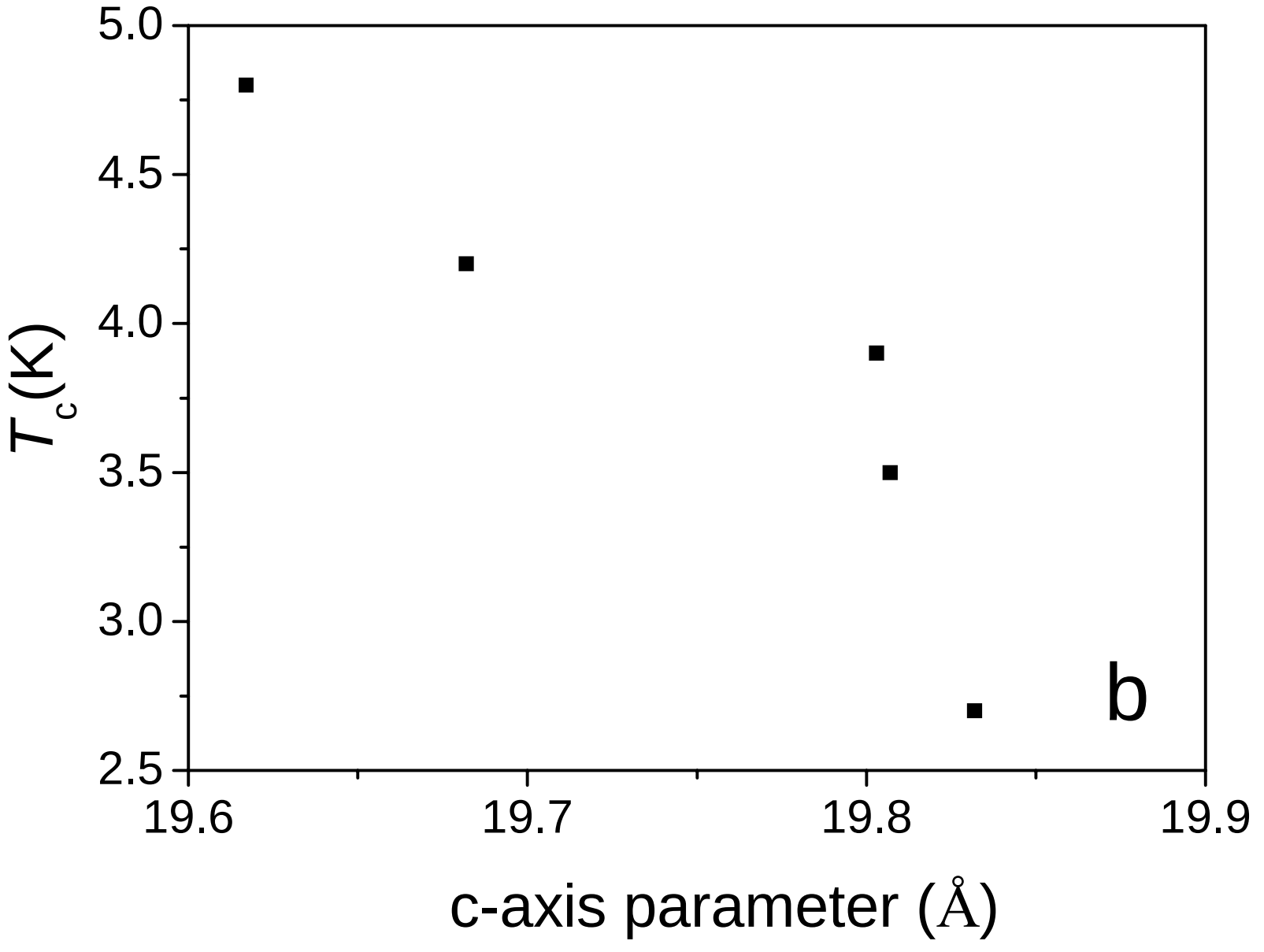

**Shi et al. (Fig. 1)**

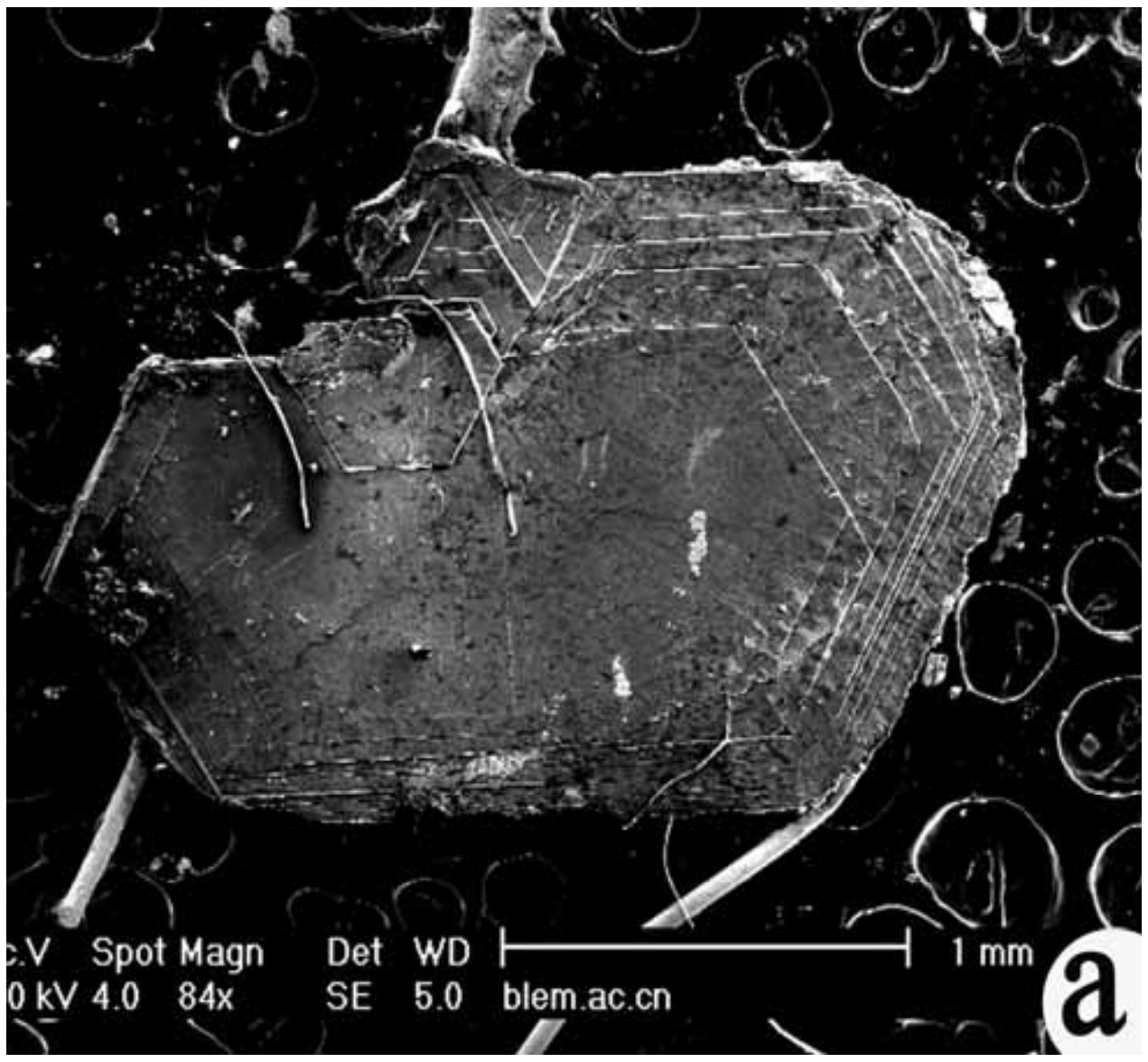

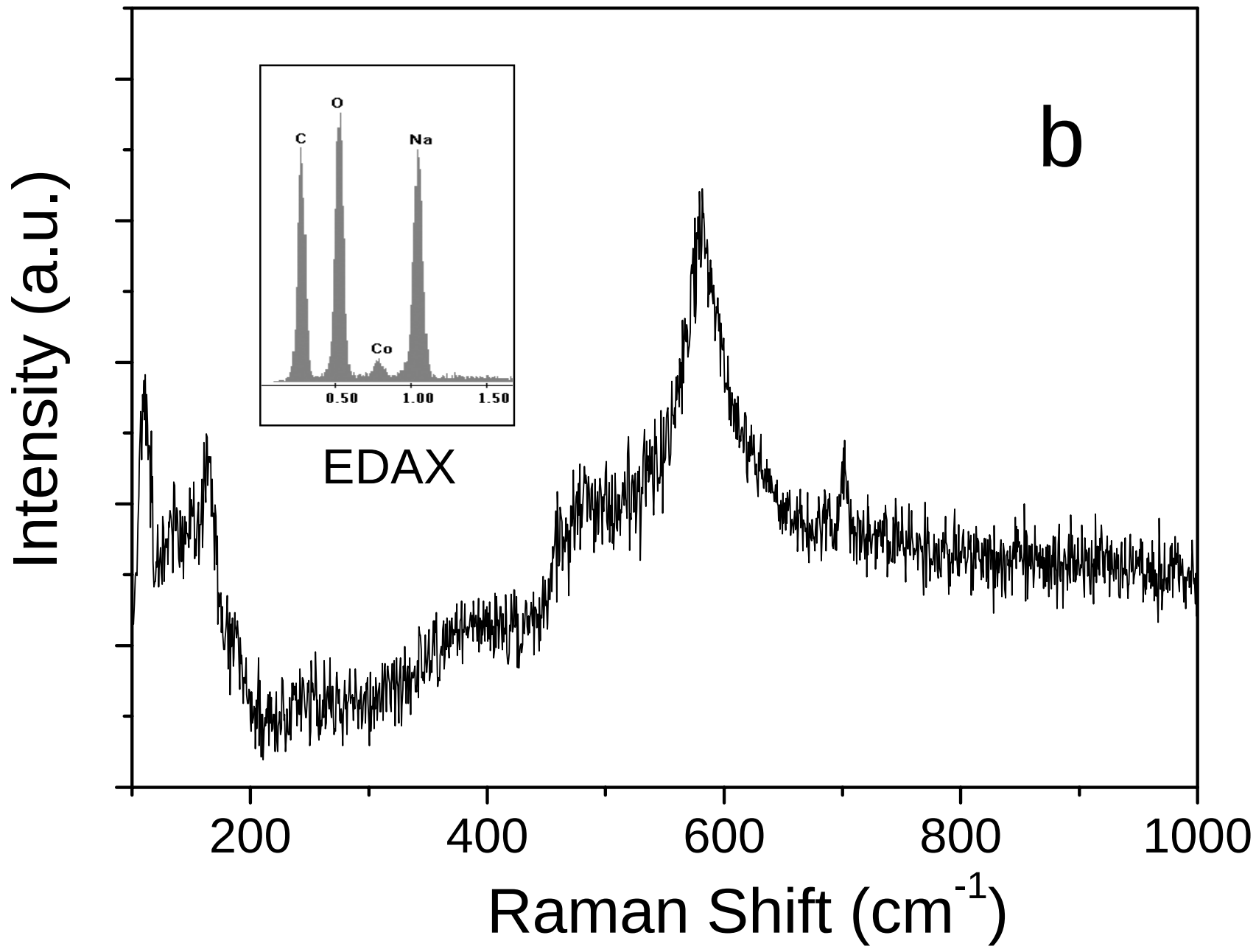

**Shi et al. (Fig. 2)**

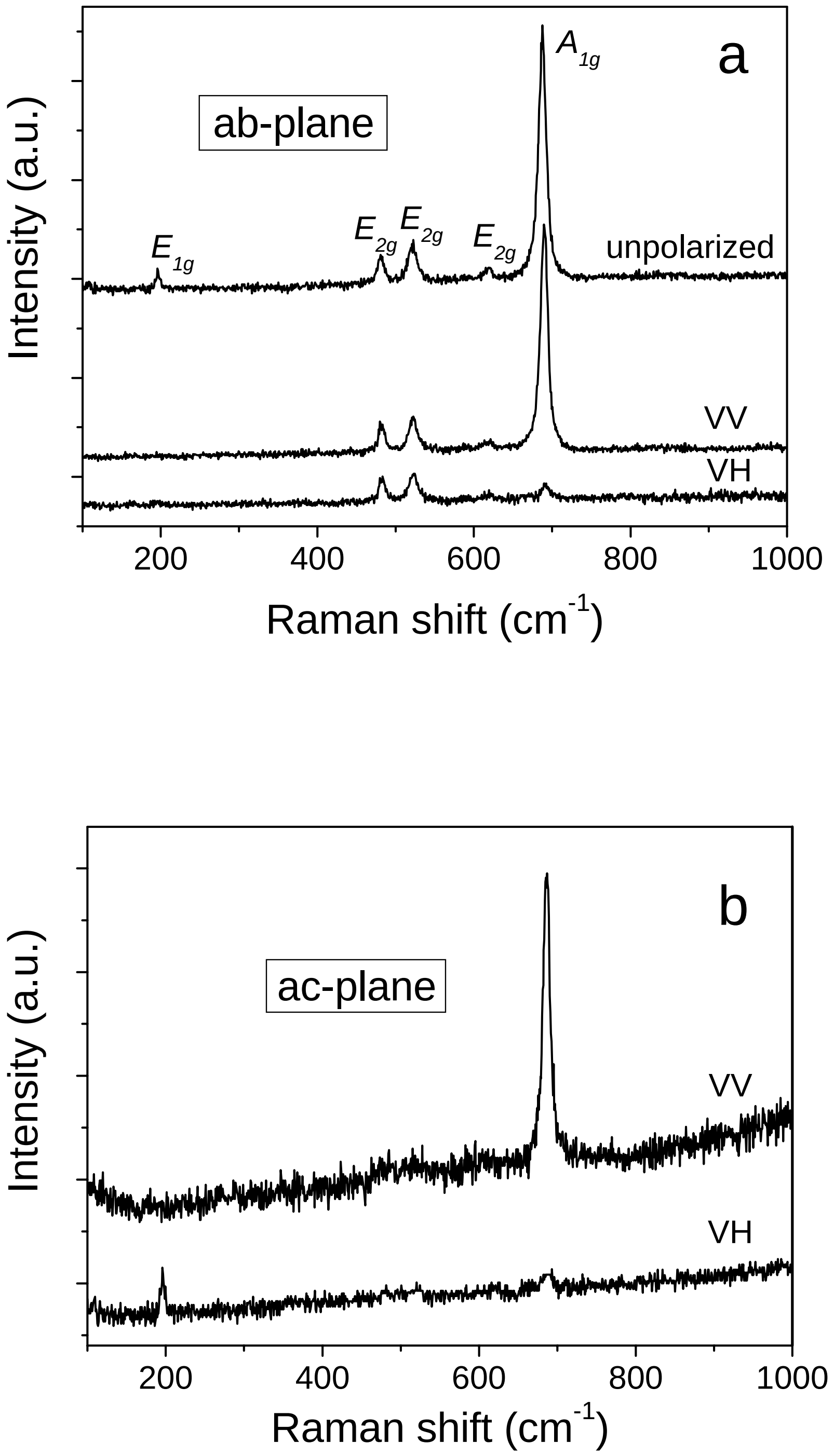

**Shi et al. (Fig. 3)**

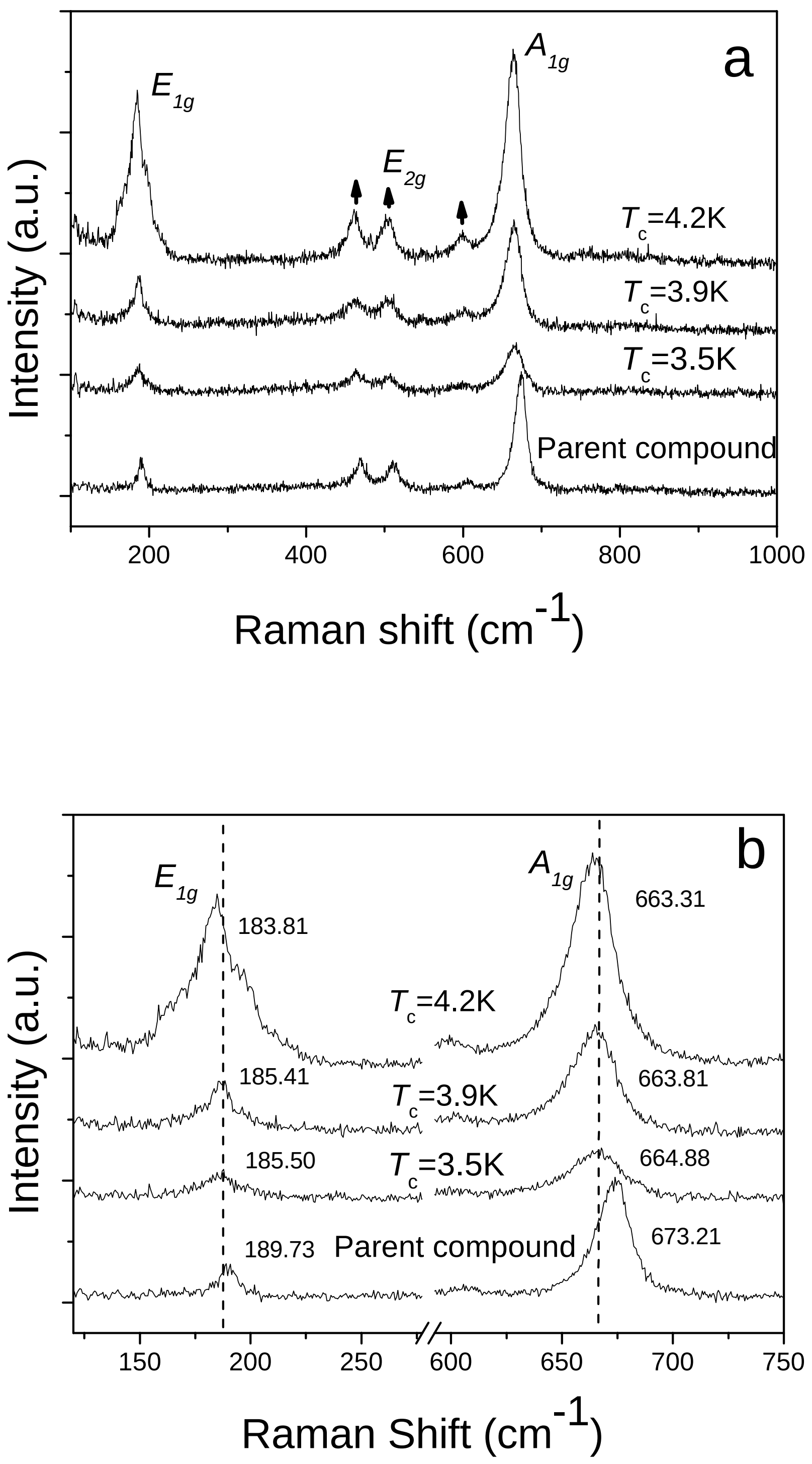

**Shi et al. (Fig. 4)**